\documentclass[3p,times]{elsarticle}

\usepackage{ecrc}

\usepackage{epstopdf}

\volume{00}

\firstpage{1}

\journalname{Nuclear Physics A}

\runauth{M. Nahrgang et al.}

\jid{nupha}

\jnltitlelogo{Nuclear Physics A}

\usepackage{graphicx}
\usepackage{amsmath,amssymb}

\begin{document}

\begin{frontmatter}

\title{Heavy-flavor observables at RHIC and LHC}

\author[1]{Marlene Nahrgang}
\author[2]{J\"org Aichelin}
\author[1]{Steffen Bass}
\author[2]{Pol Bernard Gossiaux}
\author[2]{Klaus Werner}

\address[1]{Department of Physics, Duke University, Durham, North Carolina 27708-0305, USA}
\address[2]{SUBATECH, UMR 6457, Universit\'e de Nantes, Ecole des Mines de Nantes,
IN2P3/CNRS. 4 rue Alfred Kastler, 44307 Nantes cedex 3, France}

\begin{abstract}
We investigate the charm-quark propagation in the QGP media produced in ultrarelativistic heavy-ion collisions at RHIC and the LHC. Purely collisional and radiative processes lead to a significant suppression of final $D$-meson spectra at high transverse momentum and a finite flow of heavy quarks inside the fluid dynamical evolution of the light partons. The $D$-meson nuclear modification factor and the elliptic flow are studied at two collision energies.
We further propose to measure the triangular flow of $D$ mesons, which we find to be nonzero in non-central collisions.
\end{abstract}


\end{frontmatter}


\section{Introduction}
\label{intro}
At RHIC and the LHC the Quark-gluon plasma (QGP) is produced in ultrarelativistic heavy-ion collisions. The soft sector of light partons is assumed to achieve local thermal equilibrium within the first tenth of a fm/c and the subsequent evolution of the bulk matter can be described very well by fluid dynamics.  The hard part of the transverse momentum, $p_T$, spectra and heavy quarks, like charm and bottom, are predominantly produced in the initial hard scatterings of partons in the incoming nuclei. Due to the large mass and the fast expansion heavy quarks are believed to not thermalize fully in the QGP medium and thus carry information about the interaction processes with the constituents of the QGP. The experimental results of the nuclear modification factor $R_{\rm AA}$ and the elliptic flow $v_2$ \cite{rhic,Adamczyk:2014uip, Alice,delValle:2012qw,Abelev:2014ipa} suggest a significant medium modification of the heavy-meson spectra as compared to proton-proton collisions: a high-$p_T$ suppression and a finite $v_2$ comparable in size to that of light hadrons.

The gross features of the in-medium energy loss can be described by elastic scatterings  \cite{elastic}, gluon bremsstrahlung \cite{radiative} or a combination of both. It remains a challenge to simultaneously describe the $R_{\rm AA}$ and $v_2$ data within numerical simulations, where the heavy-quark propagation is coupled to a fluid dynamical description of the medium \cite{results,Gossiaux:2008jv,Nahrgang:2013xaa} or a parton cascade \cite{bamps}.

In order to enhance our understanding of the heavy-flavor dynamics in the QGP it becomes necessary to study the centrality and mass dependence of heavy-flavor observables and to explore the potential of novel observables, such as azimuthal correlations \cite{Nahrgang:2013saa} and higher-order flow coefficients as proposed in this work.
We first introduce the model MC@sHQ+EPOS, which was developed by some of the authors, in section \ref{sec:model}. First results and details of this model can be found in \cite{Nahrgang:2013xaa,Nahrgang:2013saa}. It couples a Monte Carlo propagation of heavy quarks to the $3+1$ dimensional fluid dynamical evolution of the QGP from EPOS initial conditions  \cite{EPOS} and reproduces well the measured observables at LHC and RHIC as discussed in section \ref{sec:results}. With the same parameters we then calculate the triangular flow of $D$ mesons in non-central collisions at RHIC and the LHC.

\section{MC@sHQ+EPOS}\label{sec:model}
This model describes the heavy-quark production in heavy-ion collisions via the following three stages: 1) initialization of heavy-quark pairs according to the momentum distribution from FONLL \cite{fonll} and the spatial distribution of the nucleon-nucleon collision points, 2) in-medium propagation by an explicit Monte Carlo simulation of the Boltzmann equation, and 3) hadronization of heavy quarks into heavy mesons via coalescence and fragmentation on the hypersurface of constant temperature $T=155$~MeV.

The interaction of the heavy quarks with the thermal medium constituents can either be purely collisional or include the emission of bremsstrahlung gluons. The elastic cross sections are obtained from pQCD calculations including a running coupling $\alpha_s$. The gluon propagator in the $t$-channel is regularized within the HTL+semihard approach \cite{Gossiaux:2008jv} by $1/t\to1/(t-\kappa m_D^2(T))$ for a self-consistently determined Debye mass $m_D$. The parameter $\kappa\sim0.2$ is obtained from the requirement that the average energy loss calculated in HTL is reproduced with a maximal insensitivity of the intermediate scale between nominally soft and hard processes.
The incoherent radiation spectra result from extending the calculation in \cite{Gunion:1981qs} to finite quark masses. The effect of coherence is implemented by an effective reduction of the power spectrum per unit length \cite{Gossiaux:2012cv}. It turns out that this LPM suppression becomes important only at sufficiently high $p_T$. 
The resulting scattering rates must be understood in connection with the underlying medium description and the interpretation of the equation of state in terms of the dynamical degrees of freedom \cite{Nahrgang:2013xaa}. 
Due to the existing uncertainties in implementing the different ingredients we calibrate the energy-loss models globally through rescaling of rates by a $K$-factor. It is chosen to best reproduce the high-$p_T$ $D$ meson $R_{\rm AA}$ data in central collisions and remains unchanged when studying further observables. $K^{\rm coll+rad}=0.8$ for the collisional+radiative(LPM) scenario is rather close to a generic value of $K=1.0$.

For the background medium we choose the $3+1$ dimensional ideal fluid dynamical evolution stemming from EPOS initial conditions, which combine pQCD calculations of the hard scattering with the Gribov-Regge theory of the phenomenological, soft part of the interaction. Multiple scatterings form parton ladders, which are represented by flux tubes. Jet components can be subtracted and the bulk matter is mapped onto the initial fluid dynamical fields. The flux tube radii are enhanced in order to mimic viscous effects. The EPOS model describes well $p_T$ spectra, flow harmonics and correlation patterns like the ridge in Au+Au collisions at RHIC and Pb+Pb, p+Pb and p+p collisions at the LHC \cite{EPOS}. 
For illustration purpose, we apply a simple convolution of the initial $p_T$ spectrum in order to include (anti-)shadowing at (high) low $p_T$ in central collisions at the LHC according to the EPS09 nuclear shadowing effect \cite{Eskola:2009uj}. 
In an upcoming next EPOS version shadowing and viscous corrections will be included consistently. 

The EPOS fluctuating initial conditions allow us to perform an event-by-event study of heavy-quark observables. We optimize computing resources by running  $N^{\rm HQ}=10^4$ heavy-quark events per EPOS fluid dynamical event. It turns out that further decreasing $N^{\rm HQ}$ does not change the results significantly.

\section{Nuclear modification factor and flow}\label{sec:results}

\begin{figure}
 \begin{center}
 \includegraphics[width=0.46\textwidth]{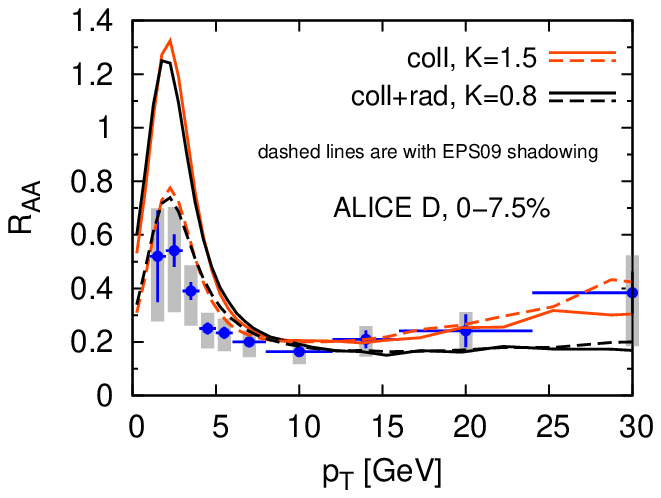}\hfil
 \includegraphics[width=0.46\textwidth]{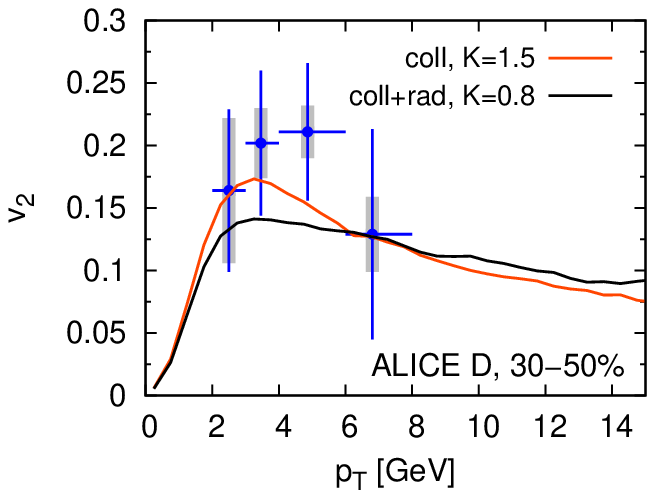}
 \caption{$D$ meson $R_{\rm AA}$ and elliptic flow $v_2$ for the centrality classes as stated in Pb+Pb collisions at $\sqrt{s_{\rm NN}}=2.76$~TeV. Both models, purely collisional (light) and collisional+radiative(LPM) (dark) are shown. Experimental data from \cite{delValle:2012qw,Abelev:2014ipa}. The last two data points for $v_2$ are outside the range of the plot.}
 \label{fig:LHC}
 \end{center}
 \end{figure}

\begin{figure}
 \begin{center}
 \includegraphics[width=0.46\textwidth]{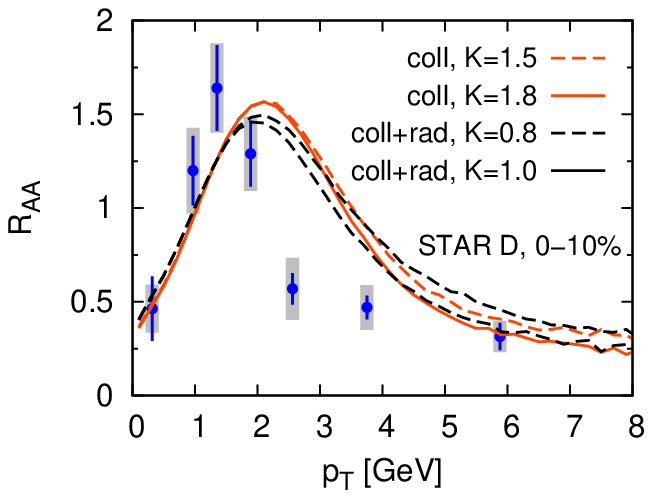}\hfil
 \includegraphics[width=0.46\textwidth]{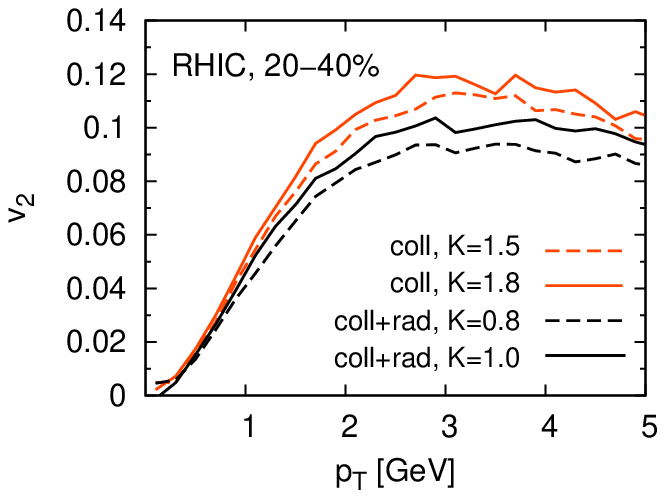}
 \caption{$D$ meson $R_{\rm AA}$ and elliptic flow $v_2$ for the centrality classes as stated in Au+Au collisions at $\sqrt{s_{\rm NN}}=200$~GeV. Both models, purely collisional (light) and collisional+radiative(LPM) (dark) are shown for two different $K$-factors, see text for details. Experimental data from \cite{Adamczyk:2014uip}.}
 \label{fig:RHIC}
 \end{center}
 \end{figure}

Figs.~\ref{fig:LHC} and \ref{fig:RHIC} present our results for the traditional heavy-flavor observables, $R_{\rm AA}$ and $v_2$, at LHC and RHIC energies. 
While the purely collisional energy loss shows the rising trend at larger $p_T$ seen in the LHC $R_{\rm AA}$ data, it is missed by the collisional+radiative(LPM) energy loss model. At low $p_T$ the data is overestimated if the initial state shadowing is not included. At higher $p_T$ an indication for the antishadowing can be seen within the current accuracy. Both energy loss models produce a substantial elliptic flow coefficient $v_2$ in non-central collisions. At lower $p_T$ the $v_2$ in the purely collisional interaction is larger than the one in the collisional+radiative(LPM) case, while at higher $p_T$ this picture is reversed. Here, energy loss is the dominant mechanism to produce an azimuthal anisotropy. We checked that as expected initial state shadowing does not affect the $v_2$ in the presently achieved accuracy.

The current range in $p_T$ of the $D$ meson $R_{\rm AA}$ measured at RHIC falls in the region where already at the LHC we are currently unable to describe the data satisfactorily without shadowing. In contrast to the LHC, however, the pattern is slightly different and the data at $p_T\sim1-1.5$~GeV is underpredicted. Applying the same $K$-factors as for the LHC, we find that the data point at highest available $p_T\sim6$~GeV is missed. We slightly adjust the $K$-factor to reproduce this data point. The consequences of this change are seen in the difference of the $R_{\rm AA}$ curves as well as in the $D$ meson elliptic flow $v_2$, where the larger $K$-factors leads to a larger $v_2$. It has not yet been measured at RHIC, except for minimum bias events. We find the same ordering from the purely collisional to the collisional+radiative(LPM) interaction mechanism as at the LHC in the given $p_T$ range, but the peak value is reduced by $\sim\!30$\%.

\begin{figure}
 \begin{center}
 \includegraphics[width=0.46\textwidth]{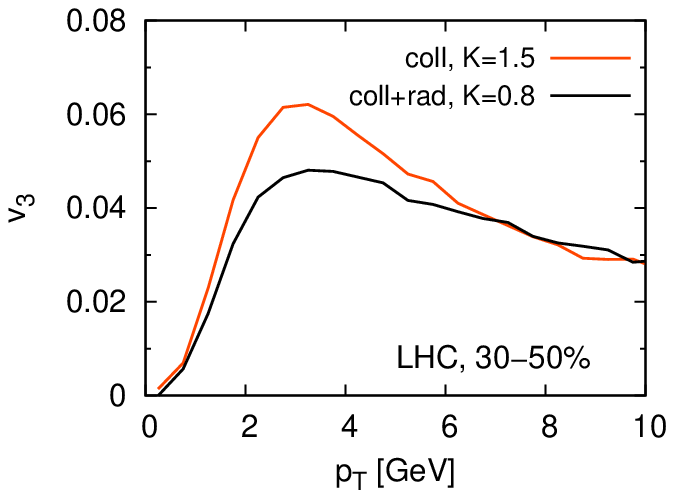}\hfil
 \includegraphics[width=0.46\textwidth]{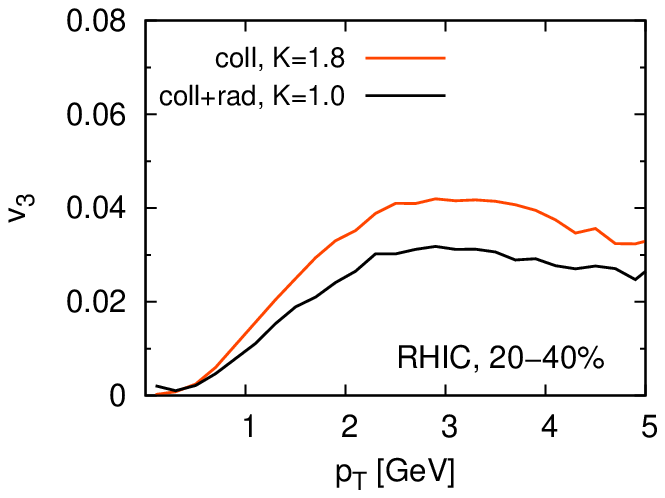}
 \caption{$D$ meson triangular flow for both interaction models and both collision systems.}
 \label{fig:v3}
 \end{center}
 \end{figure}

We finally present the first calculations of the triangular flow coefficient $v_3$ of $D$ mesons (see Fig.~\ref{fig:v3}). This is possible due the fluctuating EPOS initial conditions, where finite triangularities in the initial geometry are produced. 
In response, the $D$ mesons develop a nonzero $v_3$, which has been calculated in correlation with the initial participant plane angle $\psi_3$. It is smaller than the $v_2$ and shows the expected ordering with respect to the different energy loss mechanisms within the given accuracy.

\section{Conclusions}
We presented the results for the traditional heavy-flavor observables, $R_{\rm AA}$ and $v_2$, for RHIC and LHC energies obtained in the MC@sHQ+EPOS model. The importance of including initial state shadowing was demonstrated in the case of central collisions at the LHC. 
In the light hadron sector the measurement of higher-order flow coefficients has helped enormously in constraining the shear viscosity of the QGP and the initial conditions \cite{hydroflow}. We are proposing that measuring a nonzero $v_3$ in the heavy-quark sector can further support the substantial in-medium modifications and possibly help determining the transport properties.

\section*{Acknowledgements}
M.N. was supported by a fellowship within the Postdoc-Program of the German Academic Exchange Service (DAAD). This work was supported by the U.S. department of Energy under grant DE-FG02-05ER41367. We are grateful for support from  the Hessian LOEWE initiative Helmholtz International 
Center for FAIR, ANR research program ``hadrons @ LHC''  (grant ANR-08-BLAN-0093-02),  TOGETHER project R\'egion Pays de la Loire and I3-HP.

\end{document}